# Unified Analytical Modeling of the Error Rates and the Ergodic Channel Capacity in $\eta-\mu$ Generalized Fading Channels with Integer $\mu$ and Maximal Ratio Combining Receiver

Ehab Salahat, *Member*, *IEEE*, and Murad Qasaimeh, *Member*, *IEEE*

*Abstract*—In this paper, we introduce a novel performance analysis of the $\eta-\mu$ generalized radio fading channels with integer value of the $\mu$ fading parameter, i.e. with even number of multipath clusters. This fading model includes other fading models as special cases such as the Nakagami-*m*, the Hoyt, and the Rayleigh. We obtain novel unified and generic simple closed-form expressions for the average bit error rates and ergodic channel capacity in the additive white generalized Gaussian noise (AWGGN), which includes the additive Gaussian, the gamma, the Laplacian, and the impulsive noise as special cases. The receiver is assumed to be an *L*-branch maximal ratio combiner where we study the effect of having more deployed receiver antenna. Numerical evaluation as well as results from technical wireless literature validate the generality and the accuracy of the derived unified expressions under the studied test cases.

## I. INTRODUCTION

WIRELESS signal propagation is subjected to many uncontrollable effects that significantly degrade the performance of wireless communication systems. Such effects include the additive noise, the shadowing, and the multipath fading [1] [2] [3]. As such, it is essential to correctly and accurately model and characterize the overall wireless channel between the transmitter and the receiver, and this is where statistical channel models come into the play. Many statistical models exist that describe the characteristics of the radio signal. The Rayleigh, Rician, Weibull, Nakagami-*m*, Hoyt (Nakagami-*q*) are some of the well-known models that describe the short-term variations. Yet, some cases, none of these models seem to accurately fit measured data [4]. For example, some researchers questioned the use of the famous widely used Nakagami–*m* distribution as it seems not to fit (especial at its tail) adequately experimental measurements [5]. Similarly, the typical assumption used in the wireless research literature is that the noise is Gaussian distributed. Yet, there are few cases in which the noise in non-Gaussian (e.g. Laplacian). In this context, the need for robust wireless fading models became more clear. These models have to be flexible to provide better fitting capability and physical understanding of the wireless propagation phenomena.

The $\eta$-$\mu$ distribution [6] is a relatively new generalized fading model that gained a considerable attention in the study of wireless systems. The model is very suitable for non-line-of-sight propagation. This model is unique due to the remarkable flexibility and strong physical background. The usefulness of this model can be also observed by noting that they include, as particular cases, most of the widely used wireless fading distributions such as the Rayleigh and Hoyt.

The $\eta$-$\mu$ generalized fading model follows multi-cluster signal analysis and has two formats. In one format, the quadrature (Q) and in-phase (I) signal components are assumed to be independent with unequal powers. In the other format, also known as the $\lambda$–$\mu$ distribution, the signal's components are correlated and with similar powers. Interestingly, it is shown in [7] that the $\eta-\mu$ model can efficiently approximate the sum of independent non-identical Hoyt envelopes. The $\eta-\mu$ model can approximate the sum of independent and not identically distributed (i.n.d.) $\eta-\mu$ [8]. Likewise, the sum of independent and identical $\eta-\mu$ random variables (RV) is another $\eta$-$\mu$ RV [6] [9]. It was also shown that the sum of i.n.d. $\eta-\mu$ power variables is the sum of gamma variables with properly chosen parameters (see [10]). Some different forms of the probability density function of this fading model were also obtained in [11]. This fading was also reported to be a special case of the $\kappa-\mu$ shadowed fading model [12] [13].

In the realm of wireless performance metrics, this distribution was analyzed very recently. For example, statistics like the level-crossing rates and fading amount were analyzed in [14]. The ergodic channel capacity for this fading channel using 1-antenna at the receiving node was derived in [11] and [15]. The moment generating function (MGF) for this generalized model was obtained in both [16] and [17], where in the latter the average bit error rates (ABER) for BPSK were evaluated numerically. The ABER using multi-branch maximal ratio spatial diversity reception was also derived in [9] and [18]. Most importantly, most of the wireless performance metrics in the literature are analyzed assuming an additive Gaussian noise (AWGN), and as an entirely different and independent problem from the ACC evaluation.

To this end, in this paper, we utilize the new PDF form of the $\eta$–$\mu$ fading channels with integer value of $\mu$ (i.e. even number of multipath clusters) to obtain new closed-form expressions for average bit error rates (ABER) using spatial maximal ratio combining diversity as well as the average (ergodic) channel capacity. ABER analysis assumes the additive generalized Gaussian noise as this maximizes the generality of the derived results and investigates in details an understudied noise model. Specifically, the derivation of both performance metrics is performed using the exponential approximations of the generalized $Q$–function and the $\log_2(\cdot)$ function reported in [1]. The expressions are obtained in a unified form and in terms of the well-known hypergeometric function. Using our new derived analytical expressions, the system performance over many well-known fading models can be analyzed as well due to their accurate and generic nature.

The rest of the paper is presented as follows. The statistical analysis of the MRC output instantaneous SNR and assuming i.i.d. fading with integer $\mu$ is given in section II. The unified ABER and ACC analytical expression is derived in section III. Illustrative numerical evaluation for different fading scenarios are given in section IV. The contributions of the paper are finally briefed in section V.

## II. STATISTICAL ANALYSIS

### A. The $\eta - \mu$ Fading Distribution

The probability density function (PDF) of the instantaneous signal-to-noise ratio of the $\eta - \mu$ is given by [6] [11] [17]:

$$f_\gamma(\gamma) = \frac{2\sqrt{\pi}\mu^{\mu+0.5}h^\mu}{\Gamma(\mu)H^{\mu-0.5}\tilde{\gamma}^{\mu+0.5}}\gamma^{\mu-0.5}e^{-\left[\frac{2\mu h}{\tilde{\gamma}}\right]\gamma}I_{\mu-0.5}\left(\left[\frac{2\mu H}{\tilde{\gamma}}\right]\gamma\right), \quad (1)$$

where $I_v(\cdot)$ is the well-known modified Bessel function of the first kind [19] [20]. The parameters $h$ and $H$ given in terms of the fading parameter $\eta$. The $\eta$-$\mu$ model has two formats as shown in table I. Specifically, in format I, $\eta \in (0, \infty)$ and it signifies the power ratio between the I and Q scattered waves of the fading signal within each cluster. In format II (a.k.a. $\lambda$-$\mu$ fading), $\eta \in (-1, +1)$ and represents the correlation between the I and Q components within each cluster. In both cases, $\mu$ represents the number of multipath clusters. The $\eta$-$\mu$ model is very useful for non-line-of-sight propagation scenarios and applications. Moreover, it encloses the Rayleigh, Nakagami-$m$, Hoyt (Nakagami-$q$), and one sided Gaussian distributions as particular cases [6].

TABLE I: $h$ AND $H$ FOR $\eta - \mu$ FORMATS

| Format | $h$ | $H$ | $\eta$ |
|---|---|---|---|
| I | $(\eta^{-1} + \eta + 2)/4$ | $(\eta^{-1} - \eta)/4$ | $0 < \eta < \infty$ |
| II | $(1-\eta^2)^{-1}$ | $\eta(1-\eta^2)^{-1}$ | $-1 < \eta < 1$ |

### B. Maximal Ratio Combiner Analysis

Assuming that the wireless signal is propagated over $L$ i.i.d. $\eta - \mu$ fading channels and that MRC spatial diversity is used, the corresponding instantaneous PDF of the receiver's output SNR is given by [9]:

$$f_{MRC}(\gamma) = \frac{2\sqrt{\pi}\tilde{\mu}^{\tilde{\mu}+0.5}h^{\tilde{\mu}}}{\Gamma(\tilde{\mu})H^{\tilde{\mu}-0.5}\tilde{\zeta}^{\tilde{\mu}+0.5}}\gamma^{\tilde{\mu}-0.5}e^{-\left[\frac{2\tilde{\mu}h}{\tilde{\zeta}}\right]\gamma}I_{\tilde{\mu}-0.5}\left(\left[\frac{2\tilde{\mu}H}{\tilde{\zeta}}\right]\gamma\right), \quad (3)$$

with $\tilde{\mu} = L\mu$, and $\tilde{\zeta} = L\tilde{\gamma}$, and $L$ is the number of antennas used by the MRC receiver.

***Proof:*** the proof was presented in [9] but is also repeated here for convenience. Following the former assumptions on the propagating signal, the instantaneous SNR of the MRC output is given by [21] [22]:

$$\gamma = \sum_{i=1}^{L}\gamma_i. \quad (4)$$

Given that the MGF of the $\eta$-$\mu$ is given as [17]:

$$M_\gamma(s) = \left[\frac{4\mu^2 h}{2[(h-H)\mu+s\tilde{\gamma}][2(h+H)\mu+s\tilde{\gamma}]}\right]^\mu, \quad (5)$$

and from the i.i.d. fading assumption, the average SNR for all $L$ branches are equal, i.e. $\tilde{\gamma} = \tilde{\gamma}_1 = \tilde{\gamma}_2 = \cdots = \tilde{\gamma}_N$. As such, the MGF of the MRC output is then given as:

$$M_{\gamma_{MRC}}(s) = \prod_{i=1}^{L}M_\gamma(s) = \left[\frac{4\mu^2 h}{2[(h-H)\mu+s\tilde{\gamma}][2(h+H)\mu+s\tilde{\gamma}]}\right]^{\mu L}, \quad (6)$$

If we represent $\tilde{\zeta} = L\tilde{\gamma}$ and $L\mu$ as $\tilde{\mu}$, or $\mu = \tilde{\mu}/L$, (6) becomes:

$$M_{MRC}(s) = \left[\frac{4\tilde{\mu}^2 h}{2[(h-H)\tilde{\mu}+s\tilde{\zeta}][2(h+H)\tilde{\mu}+s\tilde{\zeta}]}\right]^{\tilde{\mu}}. \quad (7)$$

Comparing (5) and (7), one can concludes that the PDF of the MRC's output SNR with $L$ antennas is written as:

$$f_{MRC}(\gamma) = \frac{2\sqrt{\pi}\tilde{\mu}^{\tilde{\mu}+0.5}h^{\tilde{\mu}}}{\Gamma(\tilde{\mu})H^{\tilde{\mu}-0.5}\tilde{\zeta}^{\tilde{\mu}+0.5}}\gamma^{\tilde{\mu}-0.5}e^{-\left[\frac{2\tilde{\mu}h}{\tilde{\zeta}}\right]\gamma}I_{\tilde{\mu}-0.5}\left(\left[\frac{2\tilde{\mu}H}{\tilde{\zeta}}\right]\gamma\right). \quad (8)$$

This new generic PDF for this generalized fading distribution can be mapped to any of the presented equivalent PDF forms in [11] and their special cases.

For integer $\mu$, the corresponding PDF is given in [11] by:

$$f_{MRC}(\gamma) = \sum_{k=0}^{\tilde{\mu}-1}\left[\left[\frac{\tilde{\mu}}{\Omega_2-\Omega_1}\right]^{\tilde{\mu}}\left[\frac{1}{(\tilde{\mu}-1)!}\right]^2\binom{\tilde{\mu}-1}{k}\left(-\frac{1}{p}\right)^k\right] \times$$
$$\gamma^{[\tilde{\mu}-k]-1}e^{-\left[\frac{\tilde{\mu}}{\Omega_2}\right]\gamma}\mathcal{L}(k+\tilde{\mu}, p\gamma),$$
$$= \sum_{k=0}^{\mu-1}\psi\gamma^{m-1}e^{-\beta\gamma}\mathcal{L}(\xi, p\gamma), \quad (9)$$

where $\mathcal{L}(\cdot,\cdot)$ denotes the incomplete gamma function and $\psi = \left[\frac{\tilde{\mu}}{\Omega_2-\Omega_1}\right]^{\tilde{\mu}}\left[\frac{1}{(\tilde{\mu}-1)!}\right]^2\binom{\tilde{\mu}-1}{k}\left(-\frac{1}{p}\right)^k$, $m = \tilde{\mu} - k$, $\beta = \left[\frac{\tilde{\mu}}{\Omega_2}\right]$, $\xi = k + \tilde{\mu}$, $\Omega_1 = \frac{\tilde{\zeta}}{2[h+H]}$, $\Omega_2 = \frac{\tilde{\zeta}}{2[h+H]}$, and $p = \tilde{\mu}\left[\frac{1}{\Omega_1} - \frac{1}{\Omega_2}\right]$.

The PDF in (9) will be used to derive our unified and generic ABER and ACC expressions.

## III. THE UNIFIED PERFORMANCE EVALUATION

### A. ABER Analysis of MRC in $\eta - \mu$ and AWGGN

The ABER due to a given fading channels can be obtained by averaging the conditional error rates (of the noise channel) over the instantaneous fading PDF [23], that is:

$$P_e = \mathcal{A}\int_0^\infty f_\gamma(\gamma)\mathcal{Q}_a\left(\sqrt{\mathcal{B}\gamma}\right)d\gamma, \quad (10)$$

where $f_\gamma(\gamma)$ is as defined earlier, and $\mathcal{Q}_a(\sqrt{\cdot})$ signifies the AWGGN and is analytically written as [1]:

$$\mathcal{Q}_a(x) = \frac{a\Lambda_0^{2/a}}{2\Gamma(1/a)}\int_x^\infty e^{-\Lambda_0^a|u|^a}du = \frac{\Lambda_0^{2/a-1}}{2\Gamma(1/a)}\Gamma(1/a, \Lambda_0^a|x|^a), \quad (11)$$

with $\Lambda_0 = \sqrt{\Gamma(3/a)/\Gamma(1/a)}$ and $a$ being the noise parameter as given in table II, and $\mathcal{A}$ and $\mathcal{B}$ are modulation scheme dependent as given in table III.

TABLE II: VALUE OF $a$ FOR DIFFERENT NOISE MODELS.

| Noise Dist. | Impulsive | Gamma | Laplacian | Gaussian | Uniform |
|---|---|---|---|---|---|
| $a$ | 0.0 | 0.5 | 1.0 | 2.0 | $\infty$ |

TABLE III: $\mathcal{A}$ AND $\mathcal{B}$ VALUES FOR DIFFERENT MODULATIONS

| Modulation Scheme | Average SER | $\mathcal{A}$ | $\mathcal{B}$ |
|---|---|---|---|
| BFSK | $= Q_a(\sqrt{\gamma})$ | 1 | 1 |
| BPSK | $= Q_a(\sqrt{2\gamma})$ | 1 | 2 |
| QPSK, 4-QAM | $\approx 2Q_a(\sqrt{\gamma})$ | 2 | 1 |
| M-PAM | $\approx \frac{2(M-1)}{M} Q_a\left(\sqrt{\frac{6}{M^2-1}\gamma}\right)$ | $\frac{2(M-1)}{M}$ | $\frac{6}{M^2-1}$ |
| M-PSK | $\approx 2Q_a\left(\sqrt{2\sin^2\left(\frac{\pi}{M}\right)\gamma}\right)$ | 2 | $2\sin^2\left(\frac{\pi}{M}\right)$ |
| Rectangular M-QAM | $\approx \frac{4(\sqrt{M}-1)}{\sqrt{M}} Q_a\left(\sqrt{\frac{3}{M-1}\gamma}\right)$ | $\frac{4(\sqrt{M}-1)}{\sqrt{M}}$ | $\frac{3}{M-1}$ |
| Non-Rectangular M-QAM | $\approx 4 Q_a\left(\sqrt{\frac{3}{M-1}\gamma}\right)$ | 4 | $\frac{3}{M-1}$ |

Using (9), then (10) can be written as:

$$P_e = \sum_{k=0}^{\mu-1} \mathcal{A}\psi \int_0^\infty \gamma^{m-1} e^{-\beta\gamma} \mathcal{L}(\xi, p\gamma) Q_a(\sqrt{\mathcal{B}\gamma}) d\gamma, \quad (11)$$

and utilizing $Q_a(\sqrt{\cdot})$ approximation from [1], given as:

$$Q_a(\sqrt{x}) \approx \sum_{i=1}^{4} \alpha_i e^{-\lambda_i x}, \quad (12)$$

with $\alpha_i$ and $\lambda_i$ being given in Table V, then (11) can then be written in the form:

$$P_e = \sum_{i=1}^{4} \sum_{k=0}^{\mu-1} \mathcal{A}\alpha_i \psi \int_0^\infty \gamma^{m-1} e^{-\tilde{\beta}\gamma} \mathcal{L}(\xi, p\gamma) d\gamma, \quad (13)$$

with $\tilde{\beta} = [\beta + \lambda_i \mathcal{B}]$, with $\alpha_i$ and $\lambda_i$ being given in [1], which can be evaluated in a simple closed-form as:

$$P_e = \sum_{i=1}^{4} \sum_{k=0}^{\mu-1} \Psi \, _2F_1\left([\xi, m+\xi], [\xi+1], -\frac{p}{\tilde{\beta}}\right), \quad (14)$$

where $\Psi = \mathcal{A}\alpha_i \psi p^\xi \Gamma(m+\xi)[\xi \tilde{\beta}^{m+\xi}]^{-1}$ and $_2F_1([\cdot,\cdot];\cdot;\cdot)$ is the confluent hypergeometric function [19]. ABER in (14) is new and generic as it applies for all the special cases of the generalized $\eta$-$\mu$ fading and AWGGN, using MRC. In section IV, the numerical results therein are based on (14).

TABLE IV: FITTING PARAMETERS OF $Q_a(\sqrt{\cdot})$ APPROXIMATION

| $a$ | $\delta_1$ | $\delta_2$ | $\delta_3$ | $\delta_4$ | $\sigma_1$ | $\sigma_2$ | $\sigma_3$ | $\sigma_4$ |
|---|---|---|---|---|---|---|---|---|
| 0.5 | 44.920 | 126.460 | 389.400 | 96.540 | 0.130 | 2.311 | 12.52 | 0.629 |
| 1 | 0.068 | 0.202 | 0.182 | 0.255 | 0.217 | 2.185 | 0.657 | 12.640 |
| 1.5 | 0.065 | 0.149 | 0.136 | 0.125 | 0.341 | 0.712 | 10.57 | 1.945 |
| 2 | 0.099 | 0.157 | 0.124 | 0.119 | 1.981 | 0.534 | 0.852 | 10.268 |
| 2.5 | 0.126 | 1.104 | -1.125 | 0.442 | 9.395 | 0.833 | 0.994 | 1.292 |

*B. ACC Analysis in $\eta - \mu$ fading Channel*

The average (normalized) channel capacity (ACC) is one of the key performance metrics for wireless communications as it provides an upper maximum transmission rate bound in the AWGN environment. It can be analytically obtained via the averaging [21]:

$$C = \int_0^\infty \log_2(1+\gamma) f_\gamma(\gamma) d\gamma. \quad (15)$$

One can clearly notice that (10) and (15) are very similar. If the $\log_2(\cdot)$ exponential approximation given in [1] is utilized, then the $C$ will be of exactly the same form, as that in (14). Hence, $C$ is also given by:

$$C = \sum_{i=1}^{4} \sum_{k=0}^{\mu-1} \Psi \, _2F_1\left([\xi, m+\xi], [\xi+1], -\frac{p}{\tilde{\beta}}\right), \quad (16)$$

with the values of the fitting parameters being replaced accordingly. The exponential approximations of $Q_a(-)$ in (14) and $\log_2(-)$ in (16) allows the derivation of a unified ABER and ACC expression without any analytical complications.

## IV. NUMERICAL RESULTS

This section provides illustrative analytical and numerical results of the ABER in (14) and the ACC in (16). Sample test scenarios with different fading and noise conditions as well as different modulation schemes and constellation order are shown first in Fig. 1 – Fig. 4. These cases will allow us to observe and study the effect of the fading parameters $\eta$ and $\mu$ as well as the noise condition on the symbol (and hence bit) error rates. Moreover, the effect of diversity gain is also analyzed by varying the number of receiver's antenna $L$.

Initially, in Fig. 1, the analytical ABER expression in (14) is compared against the numerically evaluated ABER values, where different coherent modulation schemes with multiple constellation orders are studied in a Laplacian as well as Gaussian noise environments (see table II), using 3-antennas ($L$=3) at the MRC side. In this figure, solid lines represent the numerical plots whereas the overlaid dots correspond to (14). One can observe that the analytical and numerical results precisely agree, confirming the accuracy of (14).

As a second test, the same fading and noise conditions are assumed as those of the first test, but this time without diversity ($L$=1), following the same modulation schemes. The results, shown in Fig. 2, demonstrates an exact matching between the numerical curves and the generated results from (14). In addition, as expected, there is a performance loss (or degradation) as no spatial diversity was used.

Another two test scenarios consider the special cases of the Nakagami-$m$ fading and Hoyt fading with 3-branch MRC diversity and without diversity, assuming AWGN. The results are presented in Fig. 3 and Fig. 4, respectively (refer to [6] for the parameter mapping, for which Nakagami-$m$ is obtained as $\eta \to 0, \mu = m$, and for Hoyt $\eta = [1-q]/[1+q]$ and $\mu = 1$). One can clearly see that the approximate plots (14) agree with the numerical evaluated ones.

As for the ergodic capacity, 2 test scenarios are considered and illustrated. (16) is first used to generate ACC plots for the $\eta - \mu$ (format I) fading, but in the second scenario, (16) is used to generate the ACC (with the same values of the fading parameters) but using fading format II. The results are illustrated in Fig. 5 and 6, respectively. One can see the precise agreement as the dots generated using (16) match the numerical ACC plots. It is confirmed then from ABER and ACC figures that in (14) and (16) are indeed very accurate, unified and generic.

## V. CONCLUSION

In this paper, we provided a novel performance analysis of the $\eta-\mu$ generalized radio fading channels with integer value of the $\mu$ fading parameter, i.e. with even number of multipath clusters, which includes other fading models as special cases such as the Nakagami-*m*, the Hoyt, and the Rayleigh. New unified, generic, and simple closed-form expressions for the ergodic capacity and the average error rates using MRC spatial reception in AWGGN environment were derived. Numerical simulations proved the generality and accuracy of the derived unified expressions under the different test scenarios.

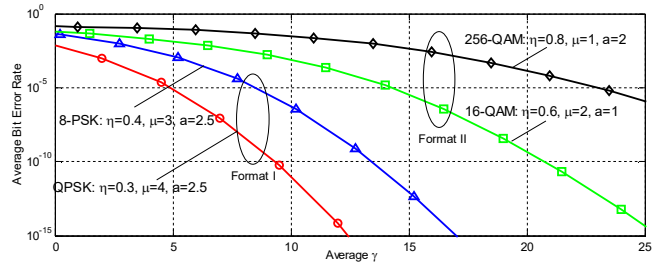

Fig. 1: Different test scenarios for the $\eta - \mu$ fading, *L*=3.

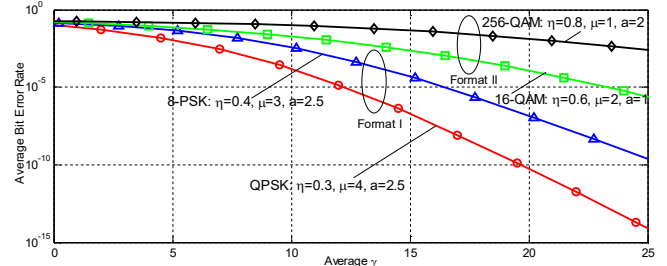

Fig. 2: Different test scenarios for the $\eta - \mu$ fading, *L*=1.

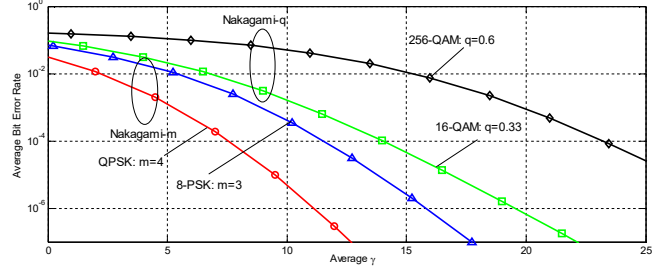

Fig. 3: Sample $\eta - \mu$ fading special cases, *L*=3 in AWGN.

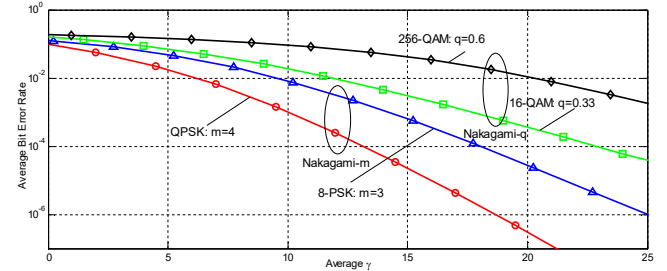

Fig. 4: Sample $\eta - \mu$ fading special cases, *L*=1 in AWGN.

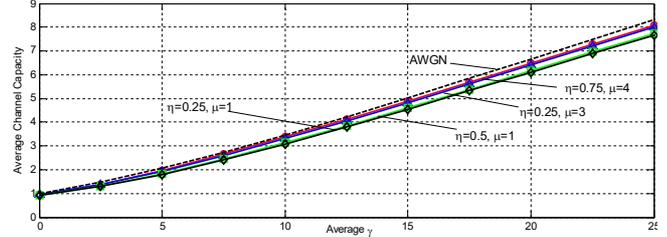

Fig. 5: Sample $\eta - \mu$ (format I) fading ACC curves.

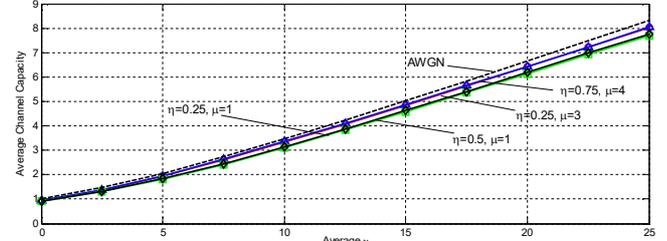

Fig. 6: Sample $\eta - \mu$ (format II) fading ACC curves.